\documentclass[preprint, review, 12pt]{article}

\usepackage[utf8]{inputenc}

\begin{document}

%


\providecommand{\keywords}[1]{\textbf{\textit{Keywords---}} #1}

\title{Paths to Unconventional Computing: Causality in Complexity}


\author{Hector Zenil\footnote{Extended version of an invited contribution to a special issue of the journal of Progress in Biophysics \& Molecular Biology (Elsevier).}}

\maketitle

\begin{abstract}
I describe my path to unconventionality in my exploration of theoretical and applied aspects of computation towards revealing the algorithmic and reprogrammable properties and capabilities of the world, in particular related to applications of algorithmic complexity in reshaping molecular biology and tackling the challenges of causality in science.
\end{abstract}


\pagebreak

The line between unconventionality, dogmatism, indeed even esotericism is very fine and critical, even in science. Turing, for example, challenged his own concept, and came up with the idea of an oracle machine to explore the implications of his challenge, though he never suggested that such a machine existed. He continued challenging conceptions with his ideas about thinking machines and processes in biology that could be closely simulated by mathematical equations, yet never suggested that machines could (or could not) think as humans do, which is why he designed a pragmatic test. Nor did he ever suggest that biology followed differential equations. Einstein, in turn, kept looking for ways to unify his gravitational and quantum models of the world, kept challenging the idea of the need for true randomness in quantum mechanics, but fell short of challenging the idea of a static (non-expanding) universe. Successful theories cannot, however, remain forever unconventional, but people can. 

My first unconventional moment, of a weak type, came when I faced the philosophical conundrum regarding the practice and the theory of computation: could the kind of mechanical description introduced by Turing be generalized not only to the way in which humans (and now digital computers) perform calculations but to the way in which the universe operates? Beyond the particularities of the Turing model, the notion of discreteness is at the heart of the digital, and it is fundamental in many ways, for example, in controlling and preventing the accumulation of errors; it is what allows us to make perfect copies of any digital data based on the work of Shannon~\cite{shannon}. It is not that I, or anyone else, ever wonders if the universe operates on a digital tape with a reading head. Rather we wonder whether the universe can in any way be reduced to a giant, high-dimensional tape as in a Turing machine, where the tape cells are discrete, and particles, for example elementary particles, interact with each other essentially like the tape cells of a digital computer. Contrary to what many may think, this is not an unconventional notion; physics points in the direction of a Turing-universe, where elementary particles cannot be further reduced in size or type. Such particles have no other particularity to them, no distinctive properties; they are exactly alike (except for its \textit{spin}), indistinguishable, just as cells on a Turing machine tape are indistinguishable except in terms of the symbol they may contain (equivalent to reading the spin direction). Moreover, classical mechanics prescribes full determinism, and the necessity of quantum mechanics to require or produce true indeterministic randomness is contested by different interpretations (e.g. Everett's multiverse).

It was not long before I started finding models of computation more powerful, at least on paper, than Turing-equivalent models. I was, however, able to find very few such models that could be serious contenders, among the many that looked exotic and thus highly unlikely or unrealizable, with some of their proponents ranging from the unconventional to the eccentric. Some propose trivial modifications of the classical model or go so far as to embrace mysticism--a strong belief that their model captures a property of the real world that no other model can~\cite{zenilcie2017}. All models beyond the Turing model involve not only infinite numbers but infinite non-computable numbers~\cite{zenil2006calcul}, that is, numbers whose digits one cannot calculate by mechanistic, Turing-type means. 

The notion that nature computes should not seem unconventional, but it apparently does. Indeed, the very fact that we have taken some material from the earth's surface and reshaped it into working electronic digital computers tells us not only that the universe can compute but that it does, and that it computes exactly as we instruct it to. The question is whether the universe carries on computations of a similar kind even without our intervention and with something other than the kind of artificial electronic computers that we build.

Every model in physics is computational and lives in the \textit{computational universe}~\cite{nks} (the universe of all possible programs), as we are able to code such models in a digital computer, plug in some data as initial conditions and run them to generate a set of possible outcomes for real physical phenomena with staggering predictive precision. We do this with ever increasing accuracy, whether calculating planetary trajectories or forecasting the weather, and such a convergence between simulation and simulated cannot but suggest the possibility that the real phenomenon performs the same or a very similar computation as the one carried out by the digital computer. We may be pushed to believe that the inadequacy of such models in predicting long term weather patterns with absolute precision reflects the limitations of the models themselves, or else the fundamental unsoundness of computable models as such, but we know that the most salient limitation here is inadequate data--both in quantitative and qualitative terms--that we can plug into the model, as we are always limited in our ability to collect data from open environments, in respect to which we can never attain enough precision without having to simulate the whole universe. But we do know that the more data we introduce into our models the better they perform. 

Computational or not, if anything was clear and not in the least unconventional, it was that the universe was algorithmic in a fundamental way, or at least that in light of successful scientific practice it seemed highly likely to be so. While this is a highly conventional point of view, many may view such a claim as being almost as strong as its mechanistic counterpart because, ultimately, in order to shift the question from computation to algorithms, one must decide what kind of computer runs the algorithms. However, after my exploration of non-computable models of computation~\cite{zenil2006calcul}, I began my exploration of what I call the algorithmic nature of the world. I wanted to study how random the world could be, and what the theory of mathematical randomness could tell us about the universe and the kinds of data that could be plugged into models, their initial conditions, and the noise attendant upon the plugging in of the data. This promised to give me a better understanding of whether it was the nature of the data on which a computational model ran that made it weaker and more limited, or whether it was only the quantity of the data that determined the limitations of computable models. And so I launched out on my strong unconventional path by introducing alternatives for measuring and applying algorithmic complexity, leading to exciting deployments of highly abstract theory in highly applied areas. The basic units of study in the theory of algorithmic complexity are sequences, and nothing epitomizes a natural sequence better than the DNA. Because most information is in the connections among genes and not the genes themselves, I defined a concept of the graph algorithmic complexity of both labelled and unlabelled graphs~\cite{zenil2014correlation, zenil2016methods}. However, this could not have been done if I had proceeded by using lossless compression as others have~\cite{cilibrasi2005clustering, vitanyi2000minimum}. Instead I used a novel approach based upon algorithmic probability~\cite{levin1974laws, solomonoff1964formal} that allowed me to circumvent some of the most serious limitations of compression algorithms. 

What I first did was to use the theory of algorithmic probability~\cite{levin1974laws, solomonoff1964formal}, a theory that elegantly reconnects computation to classical probability in a proper way through a theorem called the \textit{algorithmic coding theorem}, which for its part establishes that the most frequent outcome of a causal/deterministic system will also be the most simple in the sense of algorithmic complexity, that is, the shortest computer program that produces such an output. So we ran trillions of very small computer programs~\cite{delahaye2012numerical, soler2014calculating} to build an empirical probability distribution that approximates what in the theory of algorithmic probability is called the \textit{universal distribution}~\cite{levin1974laws}, used as the prior in all sorts of Bayesian calculations that require the framing of hypotheses about generating models or mechanisms that produce an object. Take a binary string of a million 1s. The theory tells us that the most likely mechanistic generator of 1s is not print(1 million 1s) but a recursive program that iterates over print(1) 1 million times, that is, in this case, the shortest possible program. For this sequence the most likely and highest causal generating program can easily be guessed, but for most sequences this is very difficult. However, the more difficult the more random, thus giving us some clues as to the causal content of the object.

When I started these approaches I was often discouraged, as I still sometimes am, and tempted to turn away from algorithmic complexity because `its \textit{uncomputabilty}' (the reviewers said), that there is no algorithm to run a computation in every case and expect the result of the algorithmic complexity of an object, because the computation may or may not end. But if we were scared away by uncomputability we would never code anything but trivial software. We have, however, progressed significantly in our ability to write all sorts of computer programs, in particular, incredibly powerful layers of sophisticated computer programs of which clearly we cannot control or fully understand as to e.g. know if they will ever get stuck. However, despite fear of the unknowns that uncomputability entails (e.g. the nonexistence of the perfect antivirus or the nonexistence of error-free software) software engineering does not only prevails but has changed the world. We have been too fearless of uncomputability when it comes to measuring actual algorithmic randomness. The truth is that not only are many of these challenges partially circumventable (due to e.g. their semi-computability character) but estimations have been proven to be sound and correspond to theoretical expectations~\cite{soler2013correspondence}.

Once I had the tools, methods and an unbreakable will, I wanted to know to what extent the world really ran on a universal mechanical computer, and I came up with measures of \textit{algorithmicity}~\cite{zenil2009algorithmic, zenilworld}: how much the outcome of a process resembles the way in which outcomes would be sorted according to the universal distribution, and of \textit{programmability}~\cite{zenil2015algorithmicity, zenil2014programmability, zenil2013behavioural, zenil2014nature}: how much such a process can be reprogrammed at will. The more reprogrammable, the more causal, given that a random object cannot be reprogrammed in any practical way other than by changing every possible bit. My colleagues, leading biological and cognitive labs, and I are implementing methods in which algorithmic information theory plays a central role, allowing us to steer and manipulate systems such as cells other than following traditional trial-and-error approaches. And we have looked at how the empirical universal distribution that we calculated could be plugged back into all sorts of challenges~\cite{zenil2014correlation, zenil2016methods, zenil2012computer, gauvrit2017human, gauvrit2014natural, gauvrit2014algorithmic, gauvrit2016algorithmic, gauvrit2015information} to help with the problem of data collection to generate a sound computational framework for model generation. Few, if any researchers could have foreseen that something as theoretical and uncomputable would eventually be put to these kinds of uses. 

When one takes seriously, however, the dictum that the world is algorithmic, one can begin to see seemingly unrelated natural phenomena from such a perspective and devise software engineering approaches to areas such as the study of human diseases~\cite{zenil2015causality}. Cancer is a most interesting case like other cellular and genetics diseases. Seen as a cell's computer program gone wrong, the question becomes how to reprogram cancer cells to return them to serving their original purposes, or how to hack their code in order to make them die (without tampering with the code of non-cancer cells)~\cite{zenil2015causality}.

It turns out that the world (including the natural world) may be more reprogrammable than we expected. By following a Bayesian approach to proving universal computation~\cite{zenil2016asymptotic,riedel2015cross}, we recently showed that class boundaries that seemingly determined the behaviour of computer programs could easily be transcended, and that even the simplest of programs could be reprogrammed to simulate computer programs of arbitrary complexity. This unconventional approach to universality, thinking outside the box, shows that, after the impossibility results of Turing, Chaitin or Martin-L{\"o}f, proof can no longer be at the core of some parts of theoretical computer science, and that a scientific approach based on experimental mathematics is required to answer certain questions, such as how pervasive Turing-universality is in the computational universe. We need more daring, unconventional thinkers who would stop fearing uncomputability and carry out this fruitful programme.

Niche disciplines that may seem unrelated work hand in hand in the approaches that I have introduced. Who would have thought that algorithmic randomness could be connected and make a tangible contribution to molecular biology in spite of its uncomputable character, and that algorithmic information theory would equip cognitive scientists with much needed new psychometric tools with which to test and validate long-standing~\cite{gauvrit2017human} suspicions about the inner workings of the human mind. While unconventional computing is about challenging some computational limits, the limits I challenge are those imposed by axiomatic frameworks and their quest for only mathematical proofs of ever-increasing abstraction. I rather take proofs from mature mathematical areas to seek for their meaning in disparate areas of science, thereby establishing unconventional bridges across conventional fields.

\end{document}